\documentclass[aps,pra,twocolumn]{revtex4}

\usepackage{dcolumn}
\usepackage{bm}
\usepackage{amsmath}
\usepackage{graphicx}
\usepackage{amsmath,amssymb}

\newcommand{\mean}[1]{\langle{#1}\rangle}

\newcommand{\bra}[1]{\langle{#1}|}
\newcommand{\ket}[1]{|{#1}\rangle}

\begin{document}

\preprint{APS/123-QED}

\title{
Replacing measurement-feedback with coherent-feedback for quantum state preparation
}

\author{Yoshiki Kashiwamura and Naoki Yamamoto}
\affiliation{%
Department of Applied Physics and Physico-Informatics,
Keio University, 
Hiyoshi 3-14-1, Kohoku, Yokohama 223-8522, Japan
}

\date{\today}

\begin{abstract}
The measurement-feedback is a versatile and powerful means, although its performance 
must be limited by several practical imperfections resulting from classical components. 
This paper shows that, for some typical quantum feedback control problems for state 
preparation (stabilization of a qubit or a qutrit, spin squeezing, and Fock state generation), 
the classical feedback operation can be replaced by a fully quantum one such 
that the state autonomously dissipates into the target or a state close to the target. 
The main common feature of the proposed quantum operation, which is called the coherent 
feedback, is that it is composed of the series of dispersive and dissipative couplings inspired 
by the corresponding measurement-feedback scheme. 
\end{abstract}

\maketitle


\section{Introduction}

Many quantum information systems contain {\it measurement feedback (MF)} processes 
such as teleportation and error correction \cite{Furusawa Book}. 
However, the classical components involved in such processes introduce practical 
imperfections due to detection loss, time delays in the signal processing, and the 
finite-bandwidth of actuators, which as a result severely limit the system performance. 
Thus, the following important question arises; 
Can we replace those classical components by fully-quantum systems that emulate the 
same functionalities?

\begin{figure}[htbp]
\begin{center}
\includegraphics[width=8.6cm]{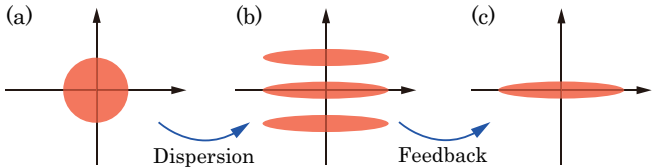}
\caption{Schematic of the feedback control for the case of squeezed state generation.
}
\label{fig:basic idea}
\end{center}
\end{figure}

The theory for MF is well established 
\cite{Belavkin,Bouten 2009,Wiseman Book,Jacobs Book,NY Book}. 
In particular, MF control method based on the quantum non-demolition (QND) measurement 
followed by the filtering (i.e., the continuous-time state estimation) has been investigated in depth 
\cite{Thomsen 2002,Handel 2005,Geremia 2006,Yanagisawa 2006,Molmer 2007,
Yamamoto 2007,Mirrahimi 2007} and some notable experiments have been demonstrated 
\cite{Haroche 2011,Siddiqi 2012,Huard 2013,Lehnert 2013,Takahashi 2013,Thompson 2016}. 
Figure~\ref{fig:basic idea} illustrates the idea of this MF control, for the case of squeezed 
state generation, as follows. 
(a) The initial state is the vacuum. 
(b) The system dispersively interacts with a probe field, and thereby they are entangled; 
if we measure the output field, the estimated system state becomes a squeezed state with 
random amplitude conditioned on the measurement result. 
The figure shows the ensemble of these conditional states. 
(c) Finally, the measurement result is fed back to compensate this random displacement for 
generating the target squeezed state deterministically.

This paper gives an answer to the question posed above. 
That is, for some typical quantum feedback control problems for state preparation, we show 
that the classical operation that compensates the random displacement (i.e., the feedback 
process in Fig.~\ref{fig:basic idea}) can be replaced by a fully quantum operation such that 
the state autonomously dissipates into the target or a state close to the target. 
Our idea is to use the {\it coherent feedback (CF)} scheme to realize this quantum operation; 
i.e., a quantum system is controlled via another quantum system in a feedback way that does 
not involve any measurement process. 
The CF scheme is implementable in a variety of systems including optics, superconductors, 
and cold atoms. 
See \cite{Wiseman 1994,Yanagisawa 2003,James 2008,Gough 2009,Nurdin 2009,
Mabuchi 2012,Yamamoto 2014,Grimsmo 2015,Jacobs 2017} for the basic theories and 
applications of CF, and 
\cite{Mabuchi 2008,Iida 2012,Devoret 2013,Kerckhoff 2013,Devoret 2016,Sarovar 2016} 
for experimental demonstrations. 
Note that the control problem considered in this paper is not contained in the framework 
where the superiority of CF over MF (or the equivalency of CF and MF) has been proved 
\cite{Wiseman 1994,Nurdin 2009,Mabuchi 2012,Yamamoto 2014,Jacobs 2017,Devoret 2016}. 
Also the proposed scheme is a sort of reservoir engineering but is different from the other 
approaches \cite{Hammerer 2004,Takeuchi 2005,Molmer 2006,Vuletic PRL 2010,Vuletic PRA 2010,Siddiqi 2012b,Lukin 2013,Vuletic 2017}, in that it relies on a novel reservoir composed 
of the series of dispersive and dissipative couplings, inspired by the MF control composed 
of the QND measurement and the subsequent filtering process.

The paper is organized as follows. 
In Sec.~II, the CF controller configuration is described in a general setting. 
Then we demonstrate how the CF can replace the MF for various state control problems: 
stabilization for a qubit (Sec.~III) and qutrit (Sec.~IV), spin squeezing (Sec.~V), and 
Fock state generation (Sec.~VI). 
Section~VII concludes the paper.


\section{The controller configuration}

\begin{figure}[htbp]
\begin{center}
\includegraphics[width=7.7cm]{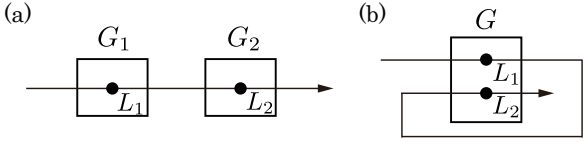}
\caption{
(a) Cascade connection of two open quantum systems $G_1$ and $G_2$. 
(b) CF for the system $G$ via cascade connection.}
\label{fig:series}
\end{center}
\end{figure}

For a general Markovian open quantum system interacting with a single probe field, the 
unconditional state obeys the master equation 
\begin{equation}
\label{eq:ME}
         \frac{d\rho}{dt} = -i[H, \rho] 
                   + L \rho L^{\dagger}
                   - \frac {1}{2}L^{\dagger}L\rho - \frac {1}{2}\rho L^{\dagger}L. 
\end{equation}
Here $L$ is the coupling operator and $H$ is a Hamiltonian; 
see Appendix~A for a detailed description of this equation. 
Thus, this system is generally characterized by $(L, H)$. 
Let us consider two open systems $G_1=(L_1,H_1)$ and $G_2=(L_2, H_2)$ 
that are unidirectionally connected through a single probe field, as shown in 
Fig.~\ref{fig:series}(a). 
Then, under the assumption that the propagation time from $G_1$ to $G_2$ is negligible, the 
whole system, denoted as $G_1\triangleright G_2$, behaves as a Markovian open system and 
is given by \cite{Gough 2009,Carmichael 1993,Gardiner Book}
\begin{equation}
\label{cascade formula}
        G_1\triangleright G_2 = \Big(L_1+L_2, H_1 + H_2 + 
              \frac{1}{2i}(L_2^\dagger L_1-L_1^\dagger L_2) \Big). 
\end{equation}

In this paper, we consider the case where $L_1, L_2, H_1$, and $H_2$ are operators 
living in the {\it same} Hilbert space associated with a single system. 
Then, as shown in Fig.~\ref{fig:series}(b), $G=G_1 \triangleright G_2$ is a CF controlled 
system where the output field after the coupling $L_1$ is again coupled to the same system 
through $L_2$. 
Moreover, $L_1$ and $L_2$ are specified as follows. 
First, $L_1$ is Hermitian; $L_1=L_1^\dagger$. 
This coupling induces a dispersive change of the system state depending on the field state. 
For the MF case, we measure the field after this coupling; 
then, ideally, the system's conditional state probabilistically changes toward one of the 
eigenstates of $L_1$, and a feedback control based on the measurement result compensates 
this randomness so that the target eigenstate is deterministically generated. 
Our CF strategy is to apply a fully-quantum dissipative process that emulates this feedback 
operation; 
that is, in Fig.~\ref{fig:series}(b), $L_2$ is chosen as a dissipative coupling operator, which 
may drive the system state to the target. 
Summarizing, the CF controlled system is given by 
\begin{eqnarray}
& & \hspace*{-2.15em}
\label{control ME}
        G=(L, H)=(e^{i\phi} L_1, H_{\rm sys}) \triangleright (L_2, 0) 
\nonumber \\ & & \hspace*{-1.1em}
           =\Big( L_2+e^{i\phi}L_1, 
               H_{\rm sys} + \frac{1}{2i}(e^{i\phi}L_2^\dagger L_1 - e^{-i\phi}L_1 L_2 ) \Big), 
\end{eqnarray}
where $L_1=L_1^\dagger$ is a given dispersive coupling and $L_2$ is a dissipative one 
to be appropriately chosen. 
Also $H_{\rm sys}$ is a system Hamiltonian and $e^{i\phi}$ represents a phase shifter 
acting on the probe field. 
In what follows we demonstrate how to choose these operators and evaluate the 
performance of the resulting CF controlled system, in some quantum control problems.


\section{Qubit stabilization}

\begin{figure}[htbp]
\begin{center}
\includegraphics[width=8.5cm]{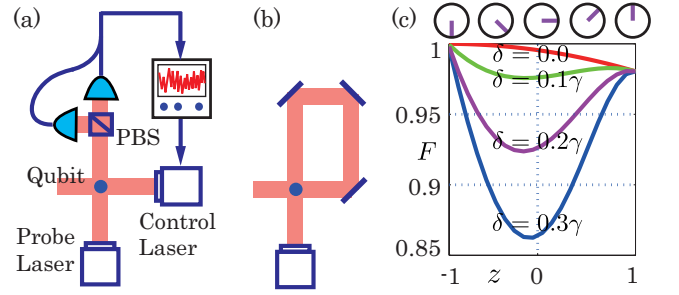}
\caption{
(a) MF and (b) CF configuration for qubit control. 
(c) Fidelity $F=\bra{\psi}\rho(\infty)\ket{\psi}$ between the ideal target state $\ket{\psi}$ 
and the steady state $\rho(\infty)$ in the realistic model. 
}
\label{fig:qubit control}
\end{center}
\end{figure}

In this section we study a qubit interacting with a probe field through the 
dispersive coupling operator $L_1=\sqrt{\kappa}\sigma_z=\sqrt{\kappa}(\ket{e}\bra{e}-\ket{g}\bra{g})$, 
where $\ket{e}=[1,\, 0]^\top$ and $\ket{g}=[0,\, 1]^\top$ 
\cite{Handel 2005,Gambetta 2008,Siddiqi 2013,Siddiqi 2014,Siddiqi 2016,Siddiqi 2017}. 
If we continuously monitor the field after this coupling, as shown in Fig.~\ref{fig:qubit control}(a), 
the qubit state conditioned on the measurement result probabilistically converges to 
$\ket{e}$ or $\ket{g}$; 
some MF control compensate this random change and realize deterministic convergence 
to $\ket{e}$ or $\ket{g}$.


\subsection{Control in the ideal setup}

First we study the CF control emulating the above MF scheme, in the ideal setting. 
Our initial task is to choose a suitable dissipative coupling $L_2$ that autonomously compensates 
the dispersive effect induced by $L_1$; 
here let us particularly take $L_2=\sqrt{\gamma}\sigma_-=\sqrt{\gamma}\ket{g}\bra{e}$, which 
represents the energy dissipation of a two-level atom with decay rate $\gamma>0$. 
Figure~\ref{fig:qubit control}(b) shows the configuration of this CF control; 
the qubit interacts with the field via $L_1$, and the output field is fed back to again couple to 
the system via $L_2$. 
Moreover we set $H_{\rm sys}=0$. 
Then the characteristic operators of this CF controlled system \eqref{control ME} are given by 
\begin{eqnarray}
       L &=&  \sqrt{\gamma}\sigma_{-} + e^{i\phi}\sqrt{\kappa}\sigma_z
            = \left[\begin{array}{cc}
                   e^{i\phi}\sqrt{\kappa} & 0  \\
                   \sqrt{\gamma} & -e^{i\phi}\sqrt{\kappa} \\ 
               \end{array} \right], 
\nonumber \\
       H &=& \frac{\sqrt{\kappa \gamma}}{2i}
                (e^{i\phi}\sigma_{-}^\dagger\sigma_{z} - e^{-i\phi}\sigma_{z}\sigma_{-}) 
\nonumber \\
\label{qubit ideal ME}
       &=& \frac{\sqrt{\kappa \gamma}}{2i}
            \left[\begin{array}{cc}
                  0 & -e^{i\phi} \\
                  e^{-i\phi} & 0 
	    \end{array}\right]. 
\end{eqnarray}
Then, noting the fact that the uniqueness of the steady state for the general finite dimensional 
master equation \eqref{eq:ME} is equivalent to the deterministic convergence to it 
\cite{Schirmer 2010}, we find that any initial state $\rho(0)$ deterministically converges to the 
following steady state $\rho(\infty)$:
\begin{equation}
\label{qubit steady state}
       \rho(\infty)=\ket{\psi}\bra{\psi},~~~
       \ket{\psi}=
          \frac{1}{\sqrt{4\kappa+\gamma}}
                \left[\begin{array}{c}
                     2e^{i\phi}\sqrt{\kappa} \\
                     \sqrt{\gamma} \\
             \end{array}\right].
\end{equation}
Interestingly, this is a {\it pure} state. 
Also, an {\it arbitrary} pure state, except $\ket{e}$, can be prepared by suitably choosing 
the control parameters $\gamma$ and $\phi$. 
$\ket{e}$ can be approximately generated by setting $\gamma \ll \kappa$, although we should 
note that the dispersive coupling is usually realized in the so-called weak coupling regime 
where $\kappa$ is relatively small. 
Recall now that the MF control can exactly stabilize $\ket{e}$ in an ideal setup, while it 
cannot stabilize any pure state other than $\ket{e}$ and $\ket{g}$. 
Hence, this CF is not a control scheme that outperforms the MF. 
Rather, the important fact we have learned through this case study is that the CF scheme 
certainly has an ability to emulate the functionality of MF, i.e., the ability to compensate 
the dispersion effect by autonomous dissipation and as a result generate a desired 
unconditional state.

Before closing this subsection, we provide another way to prove the unique convergence of 
the CF-controlled system to the state $\ket{\psi}$ given by Eq.~\eqref{qubit steady state}. 
We use the following theorem: 

{\it Theorem 1} \cite{Yamamoto 2005,Kraus 2008}: 
A pure state $\ket{\Psi}$ is a steady state of the master equation \eqref{eq:ME} if and only if 
$\ket{\Psi}$ is a common eigenvector of $L$ and $iH+L^\dagger L/2$. 

Now the eigenvectors of the operator $L$ are given by $\ket{g}$ and $\ket{\psi}$. 
Then it is immediate to see that $\ket{\psi}$ is an eigenvector of 
\[
     iH+\frac{1}{2}L^\dagger L 
      = \left[\begin{array}{cc}
             (\kappa+\gamma)/2 & -e^{i\phi}\sqrt{\kappa\gamma} \\
             0 & \kappa/2 \\
         \end{array}\right], 
\]
but $\ket{g}$ is not. 
Thus, from the above theorem, $\ket{\psi}$ is a {\it unique} steady state; 
actually, if there exists a {\it mixed} steady state, then $\ket{g}$ must also be a steady 
state due to the convexity of the Bloch sphere, which is contradiction. 
As a result, any initial state $\rho(0)$ converges to $\ket{\psi}$. 
\\

{\it Remark 1:} 
Let us consider the setup where the two couplings occur in a wrong order; 
that is, the field first couples with the system via the dissipative operator 
$L_1=\sqrt{\gamma}\sigma_-$ and secondly with the dispersive one 
$L_2=\sqrt{\kappa}\sigma_z$ in the feedback way. 
Then the operators of the CF controlled system are given by 
\begin{eqnarray}
       L &=&  \sqrt{\kappa}\sigma_z + e^{i\phi}\sqrt{\gamma}\sigma_-
            = \left[\begin{array}{cc}
                   \sqrt{\kappa} & 0  \\
                   e^{i\phi}\sqrt{\gamma} & -\sqrt{\kappa} \\ 
               \end{array} \right], 
\nonumber \\
       H &=& \frac{\sqrt{\kappa \gamma}}{2i}
                (e^{i\phi}\sigma_z \sigma_{-} - e^{-i\phi}\sigma_{-}^\dagger \sigma_z) 
       = \frac{\sqrt{\kappa \gamma}}{2i}
            \left[\begin{array}{cc}
                  0 & e^{i\phi} \\
                  -e^{-i\phi} & 0 
	    \end{array}\right]. 
\nonumber
\end{eqnarray}
In this case, the ground state $\ket{g}=[0, 1]^\top$ is the unique steady state of the master 
equation; hence any initial state converges to $\ket{g}$. 
This is a reasonable result, because what the CF controller considered here is doing 
is to emulate the operation such that the stabilizing control for $\ket{g}$ is performed 
{\it before} the measurement. 
Therefore, though not useful, this result also shows the fact that the all-quantum CF scheme 
certainly has an ability to emulate the measurement-feedback operation.


\subsection{Control performance in the imperfect setting}

To demonstrate the control performance of the proposed CF scheme in a realistic situation, 
here we consider the setup of circuit QED \cite{Gambetta 2008}; 
this paper presented a method for continuously monitoring a superconducting charge qubit 
that dispersively couples to a transmission line resonator. 
The master equation for the CF controlled qubit, which takes into account the imperfections 
studied in \cite{Gambetta 2008}, is given by 
\begin{equation}
\label{realistic ME}
      \frac{d\rho}{dt} = -i[H+H_\delta, \rho] + {\cal D}[L]\rho 
        + {\cal D}[L_{\rm ex}^{(1)}]\rho + {\cal D}[L_{\rm ex}^{(2)}]\rho, 
\end{equation}
where $H$ and $L$ are the operators in the ideal setting given in Eq.~\eqref{qubit ideal ME}. 
That is, in the practical situation, the qubit system is driven by the external Hamiltonian 
$H_\delta=\delta \sigma_z$ with $\delta$ the detuning between the qubit transition frequency 
and the driving probe frequency. 
Moreover, the system is coupled to another uncontrollable dissipative channel characterized 
by the Lindblad operator $L_{\rm ex}^{(1)}=\sqrt{\epsilon_1} \sigma_-$ and further a 
dephasing channel $L_{\rm ex}^{(2)}=\sqrt{\epsilon_2} \sigma_z$. 
In the recent experimental study \cite{Siddiqi 2017}, which has applied the theory of 
\cite{Gambetta 2008} to perform the MF control for qubit state preparation, the system 
parameters are $\kappa/2\pi=0.13$ MHz and $\epsilon_2/2\pi=0.005$ MHz; 
hence $\epsilon_2\approx 0.04\kappa$, meaning that roughly 4$\%$ loss occurs in the 
dispersive coupling process. 
We expect further progress will be made in experiments and assume $\epsilon_2= 0.01\kappa$ 
in the simulation. 
Also we set $\epsilon_1=0.01\gamma$, i.e., 1$\%$ loss in the dissipative coupling process. 
Finally $\phi=0$ is chosen for simplicity. 
Figure~3(c) shows the fidelity between the target state $\ket{\psi}$ in 
Eq.~\eqref{qubit steady state} with $\phi=0$ and the steady state $\rho(\infty)$ of the master 
equation \eqref{realistic ME}, as a function of the $z$-component of the Bloch vector 
corresponding to $\ket{\psi}$ (the target Bloch vector is depicted for several $z$ in the top 
of Fig.~3(c)). 
Note that, from the equation 
\[
     \ket{\psi}\bra{\psi}
     = \frac{1}{4\kappa + \gamma} 
        \left[\begin{array}{cc}
               4\kappa & 2\sqrt{\kappa \gamma}  \\
               2\sqrt{\kappa \gamma} & \gamma \\ 
                    \end{array} \right]
     = \frac{1}{2} 
        \left[\begin{array}{cc}
               1+z & x \\
               x & 1-z \\ 
                    \end{array} \right],
\]
we have $z=(4\kappa/\gamma-1)/(4\kappa/\gamma+1)$. 
In the ideal setting (the case $\delta=0$, $\epsilon_1=0$, and $\epsilon_2=0$), the fidelity 
takes $F(z)=\bra{\psi}\rho(\infty)\ket{\psi}=1$ for all $z$; 
that is, as proven in the previous subsection, an arbitrary pure qubit state (except $\ket{e}$) 
can be prepared by suitably choosing the system parameter $\kappa/\gamma$. 
In the practical setting, if the detuning $\delta$ is small (desirably the case $\delta=0$ in the figure), 
the fidelity monotonically decreases as $z$ increases, due to the additional decoherence 
process $L_{\rm ex}^{(1)}=\sqrt{0.01\gamma} \sigma_-$ and 
$L_{\rm ex}^{(2)}=\sqrt{0.01\kappa} \sigma_z$. 
The figure shows that, in this case, states close to the ground state can be prepared with good 
fidelity nearly $F(z)\approx 1$. 
In particular, the superposition $(\ket{g}+\ket{e})/\sqrt{2}$ can be stabilized with fidelity bigger 
than 0.99. 
On the other hand, if $\delta$ becomes large, the fidelity function takes the minimum at around 
$z=-0.1$ and decreases down to about 0.86 when $\delta=0.3\gamma$. 
It is notable, however, that even in those cases a state close to the excited state can be 
produced with fidelity $\approx 0.97$. 
Therefore, the CF scheme functions as a robust emulator for selectively producing $\ket{e}$ 
or $\ket{g}$. 
Note of course that, in order to stabilize a superposition, the detuning should be sufficiently 
suppressed.


\section{Qutrit stabilization}

Next, let us consider a qutrit such as a three-level atom, with states $\ket{1}=[1, 0, 0]^\top$, 
$\ket{2}=[0, 1, 0]^\top$, and $\ket{3}=[0, 0, 1]^\top$. 
We assume that the following dispersive coupling $L_1$ and the dissipative one $L_2$ 
can be implemented \cite{Wallraff 2010,Sorensen 2016}: 
\begin{equation*}
    L_1   = \sqrt{\kappa}\left[
                    \begin{array}{ccc}
                         1 & 0 & 0  \\
                         0 & 0 & 0  \\
                         0 & 0 & -1 \\ 
                    \end{array} \right], ~~
    L_2=\sqrt{\gamma} \left[ 
                 \begin{array}{ccc}
                      0 & 0 & 0  \\
                      1 & 0 & 0  \\
                       0 & 1 & 0 \\
                 \end{array} \right]. 
\end{equation*}
Measuring the probe after the dispersive coupling $L_1$ produces the conditional state, 
which probabilistically converges to one of the eigenstates of $L_1$, 
$\{\ket{1}, \ket{2}, \ket{3}\}$; 
a suitable MF control can compensate this dispersive change and deterministically stabilize 
an arbitrary eigenstate \cite{Yamamoto 2007,Mirrahimi 2007}. 
As for $L_2$, this induces the state change $\ket{1}\rightarrow \ket{2}\rightarrow \ket{3}$, 
i.e., a ladder-type dissipation for a three-level atom illustrated in Fig.~\ref{fig:spin1dynamics}(a). 
This dissipation is induced by the coupling of the qutrit to a single probe field $B(t)$; 
the Hamiltonian representing this instantaneous coupling is given by (see Appendix~A) 
\[
      H_{\rm int}(t+dt, t) = 
          i\sqrt{\gamma}(\ket{3}\bra{2} + \ket{2}\bra{1}) dB^\dagger (t) + {\rm h.c.}. 
\]
%


\subsection{Control in the ideal setup}

First let us set $H_{\rm sys}=0$ and $\phi=0$. 
Then the CF controlled system \eqref{control ME}, which might be implemented in a similar 
setup as Fig.~\ref{fig:qubit control}(b), is characterized by 
\begin{equation}
\label{qutrit L H}
       L = \left[\begin{array}{ccc}
                   \sqrt{\kappa} & 0 & 0 \\
                   \sqrt{\gamma} & 0 & 0 \\ 
                   0 & \sqrt{\gamma} & -\sqrt{\kappa} \\ 
               \end{array} \right], ~~
       H = \frac{i\sqrt{\kappa \gamma}}{2}
            \left[\begin{array}{ccc}
                  0 & 0 & 0 \\
                  0 & 0 & 1 \\
                  0 & -1 & 0 \\ 
	    \end{array}\right]. 
\end{equation}
The master equation has the following unique solution: 
\[
     \rho(\infty) =  
         \frac{1}{5\kappa+\gamma}
               \left[ \begin{array}{ccc}
                      0 & 0 & 0   \\
                      0 & 4\kappa & 2\sqrt{\kappa\gamma} \\
                      0 & 2\sqrt{\kappa\gamma} & \kappa+\gamma \\
                 \end{array} \right].
\]
Unlike the qubit case, this is not a pure state; 
purity is ${\rm Tr}(\rho(\infty)^2)=1-8/(5+\gamma/\kappa)^2$. 
For instance when $\gamma=3\kappa$, $\rho(\infty)$ approximates 
$\ket{\Psi_{23}}=(\ket{2}+\ket{3})/\sqrt{2}$ with fidelity 
$\bra{\Psi_{23}}\rho(\infty)\ket{\Psi_{23}}\approx 9.33$. 
However, $\rho(\infty)$ is a particular mixed state, which can stabilize neither $\ket{1}$ 
nor $\ket{2}$.

\begin{figure}[htbp]
\begin{center}
\includegraphics[width=8.6cm]{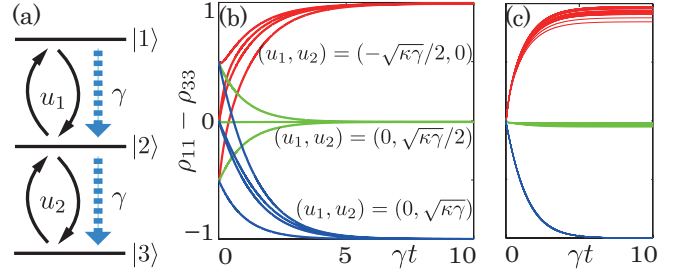}
\caption{
Energy diagram of the states (a), and the time evolution of $\rho_{11}-\rho_{33}$ in the case 
$\kappa = 100\gamma$, (b) with several initial states in the ideal setup and (c) with a specific 
initial state in the realistic setup. 
}
\label{fig:spin1dynamics}
\end{center}
\end{figure}

To emulate the MF scheme and stabilize an arbitrary eigenstate of $L_1$, the CF scheme 
needs to have a system Hamiltonian $H_{\rm sys}$ to move the steady state. 
Here we take 
\begin{equation}
\label{qutrit Hsys}
    H_{\rm sys} = iu_1 (\ket{2}\bra{1} - \ket{1}\bra{2}) 
                  + iu_2 (\ket{3}\bra{2} - \ket{2}\bra{3}), 
\end{equation}
where $(u_1, u_2)$ are real parameters to be determined; 
$H_{\rm sys}$ exchanges $\ket{1}$ and $\ket{2}$ with strength $u_1$, and 
$\ket{2}$ and $\ket{3}$ with $u_2$ as shown in Fig.~\ref{fig:spin1dynamics}(a). 
Finally we set $\phi=0$. 
Then the CF controlled system \eqref{control ME} is characterized by $L$ in 
Eq.~\eqref{qutrit L H} and 
\begin{equation}
\label{qutrit H}
       H = \left[\begin{array}{ccc}
                  0 & -iu_1 & 0 \\
                  iu_1 & 0 & -iu_2+i\sqrt{\kappa\gamma}/2 \\
                  0 & iu_2-i\sqrt{\kappa\gamma}/2 & 0 \\ 
	     \end{array}\right]. 
\end{equation}

The parameter $(u_1, u_2)$ can be determined by using Theorem~1 given in Sec.~III-A. 
Now, the eigenvectors of $L$ are calculated as 
\[
     \ket{\Phi_1} 
                = \frac{1}{2\kappa + \gamma}
                    \left[\begin{array}{c}
                         2\kappa  \\
                         2\sqrt{\kappa \gamma}  \\
                         \gamma \\ 
                    \end{array} \right], ~~
     \ket{\Phi_2} 
                = \frac{1}{\sqrt{\kappa + \gamma}}
                    \left[\begin{array}{c}
                         0  \\
                         \sqrt{\kappa}  \\
                         \sqrt{\gamma} \\ 
                    \end{array} \right], 
\]
and $\ket{\Phi_3} = \ket{3}$. 
Note that, if $\kappa\gg\gamma$, $\ket{\Phi_1}$ and $\ket{\Phi_2}$ approximates 
$\ket{1}$ and $\ket{2}$, respectively. 
Then, by solving the equation $(iH+L^\dagger L/2)\ket{\Phi_j}=\lambda_j\ket{\Phi_j}$, 
we end up with $(u_1, u_2)=(-\sqrt{\kappa\gamma}/2,0)$ for the case $\ket{\Phi_1}$, 
$(u_1, u_2)=(0, \sqrt{\kappa\gamma}/2)$ for the case $\ket{\Phi_2}$, 
and $u_2=\sqrt{\kappa\gamma}$ for the case $\ket{\Phi_3}$. 
Moreover, each $\ket{\Phi_j}$ is a {\it unique} steady state of the CF controlled system (the 
proof is given in Appendix~B), and thus any $\rho(0)$ converges to $\ket{\Phi_j}$ 
according to the result of \cite{Schirmer 2010}.

In Fig.~\ref{fig:spin1dynamics}(b) the time evolution of $\rho_{11}-\rho_{33}$ is plotted 
with several initial states $\rho(0)$ in the ideal setup. 
The parameters are taken as $\kappa=100\gamma$, hence $\ket{\Phi_1}\approx \ket{1}$ 
and $\ket{\Phi_2}\approx \ket{2}$. 
This figure shows that, by properly choosing the control parameters $(u_1, u_2)$, we can 
selectively and deterministically generate $\ket{1}$, $\ket{2}$, or $\ket{3}$. 
(Note that $\rho_{11}-\rho_{33}\rightarrow 0$ indicates $\rho\rightarrow\ket{2}\bra{2}$ 
in the figure.) 
That is, the CF scheme certainly emulates the corresponding MF control.


\subsection{Control performance in the imperfect setting}

Here we study a three-level artificial ladder-type atom implemented in a superconducting 
circuit \cite{Sorensen 2016}, as a realistic model of the qutrit system. 
The first practical imperfection is the parameter mismatch. 
Recall that we need to add the driving Hamiltonian $H_{\rm sys}$, and its parameters 
have to be exactly specified. 
For instance, if $\ket{\Phi_1}$ is the target, then the parameters must be exactly 
$(u_1, u_2)=(-\sqrt{\kappa\gamma}/2, 0)$. 
In reality, however, there exists a deviation: 
\[
      u_1=-(1+\Delta)\sqrt{\kappa\gamma}/2, 
\]
where $\Delta$ is the unknown parameter. 
Similarly, $u_2=(1+\Delta)\sqrt{\kappa\gamma}/2$ for the case of $\ket{\Phi_2}$ and 
$u_2=(1+\Delta)\sqrt{\kappa\gamma}$ for the case of $\ket{\Phi_3}$. 
The non-zero $\Delta$ would affect on the performance of control.

Next, in addition to the driving Hamiltonian $H_{\rm sys}$ given by Eq.~\eqref{qutrit Hsys}, 
the system is subjected to 
\[
        H_\delta=\delta_1 \ket{1}\bra{1} + \delta_2 \ket{2}\bra{2}, 
\]
where $\delta_1=\omega_{13}-\omega_{\rm in}-\Omega_1$ and 
$\delta_2=\omega_{23}-\omega_{\rm in}-\Omega_2$ are detunings; 
$\omega_{13}$ and $\omega_{23}$ are the transition frequency of the energy levels 
$\ket{1}\leftrightarrow\ket{3}$ and $\ket{2}\leftrightarrow\ket{3}$, respectively, with 
$\omega_{\rm in}$ the center frequency of the probe input field and $\Omega_i$ the 
frequency of the driving Hamiltonian with strength $u_i$. 
Likewise the case of parameter mismatch, the detunings also violate the condition for 
the system to have a pure steady state.

The last imperfection is decoherence. 
In addition to the ideal ladder-type decay process represented by $L_2$, in reality there 
exist independent decay processes such that the emitted photon leaks to the fields $B_1(t)$ 
and $B_2(t)$. 
This coupling is represented by the interaction Hamiltonian 
\begin{eqnarray}
& & \hspace*{0em}
     H_{\rm int}'(t+dt, t) = 
         i\sqrt{\epsilon_1}\Big( \ket{2}\bra{1} dB_1^\dagger (t) - \ket{1}\bra{2} dB_1(t) \Big)
\nonumber \\ & & \hspace*{7em}
       \mbox{}
      + i\sqrt{\epsilon_2}\Big( \ket{3}\bra{2} dB_2^\dagger (t) - \ket{2}\bra{3} dB_2(t) \Big). 
\nonumber
\end{eqnarray}
The master equation of the CF-controlled system, which takes into account the above 
imperfections, is  
\[
      \frac{d\rho}{dt} = -i[H+H_\delta, \rho]  + {\cal D}[L]\rho 
          + {\cal D}[L_{\rm ex}^{(1)}]\rho + {\cal D}[L_{\rm ex}^{(2)}]\rho, 
\]
where $L_{\rm ex}^{(1)}=\sqrt{\epsilon_1}\ket{2}\bra{1}$, 
$L_{\rm ex}^{(1)}=\sqrt{\epsilon_2}\ket{3}\bra{2}$, 
$L$ in Eq.~\eqref{qutrit L H}, and $H$ in Eq.~\eqref{qutrit H}. 
The simulation shown in Fig.~4(c) has been carried out with the following parameter 
choice. 
First we take $\kappa=100\gamma$, which realizes $\ket{\Phi_1}\approx \ket{1}$ and 
$\ket{\Phi_2}\approx \ket{2}$. 
In the ideal case where $\Delta, \delta_1, \delta_2, \epsilon_1, \epsilon_2$ are all zero, 
the qutrit state $\rho(t)$ selectively converges to one of $\{\ket{1}, \ket{2}, \ket{3}\}$, as 
demonstrated in Fig.~4(b). 
The decoherence strength is fixed to $\epsilon_1=\epsilon_2=\sqrt{\kappa\gamma}/1000$, 
in view of the fact that, in the experiment \cite{Sorensen 2016}, the corresponding parameters 
are estimated as $\epsilon=2\pi\times 0.272$ MHz and $\sqrt{\kappa\gamma}=2\pi\times 240$ MHz. 
For the detunings $(\delta_1, \delta_2)$, they take random numbers generated from the 
uniformly random distribution on $[-\sqrt{\kappa\gamma}, \sqrt{\kappa\gamma}]$. 
The parameter uncertainty $\Delta$ also takes a random number generated from the 
uniformly random distribution on $[-0.01, 0.01]$. 
The random variables $(\delta_1, \delta_2, \Delta)$ are independent. 
The simulation result with this setting is depicted in Fig.~4(c), where for each case of 
$\ket{\Phi_i}$ 30 sample paths are plotted. 
This figure clearly shows that the state convergence to $\ket{2}$ or $\ket{3}$ is robust 
against the above imperfections. 
For the case of $\ket{1}$, it looks that the fluctuation of the trajectories is not small, but 
the mean value of the fidelity $\bra{1}\rho(\infty)\ket{1}$ is 0.9531. 
Therefore, we can conclude that the CF control scheme functions as a robust state generator. 
\\

{\it Remark 2:} 
The robustness property against the detuning $H_\delta$ can be theoretically explained 
as follows, especially when $\epsilon_1=\epsilon_2=\Delta=0$. 
The iff condition for the pure state $\ket{\Phi_i}$ to be a steady state is that it is an eigenvector 
of $i(H+H_\delta)+L^\dagger L/2$; 
in the case of $\ket{\Phi_1}$, this condition is represented by 
\begin{eqnarray}
        \left[\begin{array}{ccc}
                         i\delta_1+(\kappa+\gamma)/2 & u_1 & 0 \\
                         -u_1 & i\delta_2+\gamma/2 & u_2-\sqrt{\kappa\gamma}  \\
                         0 & -u_2 & \kappa/2  \\ 
                    \end{array} \right]
          \left[\begin{array}{c}
                         2\kappa  \\
                         2\sqrt{\kappa \gamma}  \\
                         \gamma \\ 
                    \end{array} \right]
\nonumber \\
           =   \left[\begin{array}{c}
                         2i\kappa \delta_1+\kappa(\kappa+\gamma) + 2u_1\sqrt{\kappa\gamma} \\
                         -2u_1\kappa + 2i\delta_2\sqrt{\kappa\gamma} + \gamma u_2 \\
                         -2u_2 \sqrt{\kappa\gamma} + \kappa\gamma/2 \\ 
                    \end{array} \right]
           = \lambda 
           \left[\begin{array}{c}
                        2\kappa  \\
                         2\sqrt{\kappa \gamma}  \\
                         \gamma \\ 
                    \end{array} \right],
\nonumber
\end{eqnarray}
for some constant $\lambda$. 
Now we choose $(u_1, u_2)=(-\sqrt{\kappa\gamma}/2, 0)$, which are the optimal parameters 
in the ideal case $\delta_1=\delta_2=0$. 
Then, the above eigen-equation becomes 
\[
      \left[\begin{array}{c}
                         \kappa^2 + 2i\kappa \delta_1 \\
                         \kappa\sqrt{\kappa\gamma} + 2i\delta_2\sqrt{\kappa\gamma} \\
                         \kappa\gamma/2  \\ 
                    \end{array} \right]
           = \lambda 
          \left[\begin{array}{c}
                        2\kappa  \\
                         2\sqrt{\kappa \gamma}  \\
                         \gamma \\ 
                    \end{array} \right],
\]
which approximately holds with $\lambda=\kappa/2$, if $\delta_1$ and $\delta_2$ are 
much smaller than $\kappa$. 
Hence, $\ket{\Phi_1}$ is a robust steady state of the CF controlled system, under the 
influence of the detuning. 
Likewise, we can prove the robustness property of $\ket{\Phi_2}$ and $\ket{\Phi_3}$.



\section{Spin squeezing}

\begin{figure}[htbp]
\begin{center}
\includegraphics[width=5.8cm]{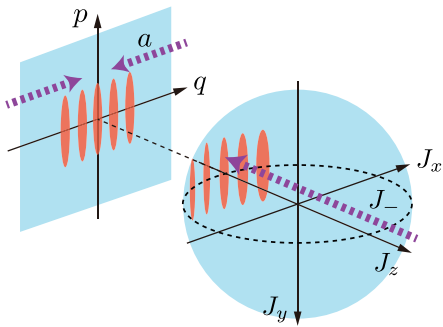}
\caption{
Projection of the spin operators to the boson opera- tors, for a large atomic ensemble.}
\label{fig:squeezing}
\end{center}
\end{figure}

We next study an atomic ensemble. 
The goal is to generate a spin-squeezed state, which can be applied for quantum 
magnetometry \cite{Nori 2011}. 
The basic variables are the spin angular momentum operators $(J_x, J_y, J_z)$. 
They satisfy $[J_x, J_y]=iJ_z$ and accordingly 
$\mean{\Delta J_x^2} \mean{\Delta J_y^2} \geq |\mean{J_z}|^2/4$, where 
$\mean{J_i}={\rm Tr}(J_i \rho)$ and $\Delta J_i = J_i - \mean{J_i}$. 
Also the lowering operator is defined as $J_-=J_x-iJ_y$. 
Here we assume that the ensemble is large, i.e., $J \gg 1$, and the state lies near 
the collective spin-down state. 
Then $J_z$ can be approximated as $J_z\approx -J$, and $(J_x, J_y)$ satisfy 
$[J_x, J_y]=-iJ$ and thus $\mean{\Delta J_x^2} \mean{\Delta J_y^2} \geq J^2/4$. 
Hence the spin operators can be transformed to the boson operators as 
$q=J_x/\sqrt{J}$, $p=-J_y/\sqrt{J}$, and $a=(q+ip)/\sqrt{2} = J_-/\sqrt{2J}$ \cite{Holstein}; 
As shown in Fig.~\ref{fig:squeezing}, this is a projection from the generalized Bloch 
sphere onto the 2 dimensional phase space.

Suppose that the atomic ensemble dispersively couples with an optical field with annihilation 
operator $B_1(t)$ via the following Faraday interaction Hamiltonian 
\cite{Thomsen 2002,Takahashi 2013,Thompson 2016}: 
\[
          H_{\rm int}^{(1)}(t+dt, t) 
               = i\sqrt{\kappa}\Big( q dB_1^\dagger (t) - q dB_1(t) \Big),
\]
meaning that $L_1=\sqrt{\kappa}q$. 
Through this interaction, the polarization of the optical probe field changes depending on 
the system's energy level. 
Hence, measuring the probe field after this coupling yields the conditional squeezed state 
with random amplitude on the $q$-axis as shown in Fig.~\ref{fig:squeezing}; 
then, as implied by Fig.~\ref{fig:basic idea}, a suitable MF can compensate this dispersive 
change and generate an unconditional squeezed vacuum state. 
Here we take the following dissipative system-field coupling, which simply represents the 
energy decay, to construct a CF that emulates this MF control: 
\[
          H_{\rm int}^{(2)}(t+dt, t) 
               = i\sqrt{\gamma}\Big( a dB_2^\dagger (t) - a^\dagger dB_2(t) \Big), 
\]
meaning that $L_2=\sqrt{\gamma}a$. 
In fact, as indicated by the purple arrows in Fig.~\ref{fig:squeezing}, this dissipative CF 
operation will stabilize a squeezed vacuum state, or equivalently a spin squeezed state 
at around $J_z\approx -J$. 
This means that an additional system Hamiltonian would not be necessary to achieve 
the goal; that is, $H_{\rm sys}=0$. 
Also we set $\phi=0$. 
Then the system operators of the CF controlled system \eqref{control ME} are given by 
\begin{eqnarray}
      H &=& \frac{1}{2i}(L_2^\dagger L_1 - L_1^\dagger L_2)
          =-\frac{\sqrt{\kappa \gamma}}{2}(qp+pq),
\nonumber \\
      L &=& L_1+L_2
         = (\sqrt{\kappa}+\sqrt{\gamma})q + i\sqrt{\gamma}p.
\nonumber
\end{eqnarray}
Note that $H\propto J_xJ_y + J_yJ_x$ is the two-axis twisting Hamiltonian \cite{Nori 2011}, 
which itself has an ability to yield a spin squeezed state. 
As noted in Sec.~I, there are several approaches for producing such a squeezing operation 
via CF \cite{Hammerer 2004,Takeuchi 2005,Molmer 2006,Vuletic PRL 2010,Vuletic PRA 2010,Lukin 2013,Vuletic 2017}, but the method proposed in this paper differs from those in that it utilizes a 
novel feedback operation composed of the series of dispersive and dissipative couplings 
inspired by the corresponding MF control.

Now, $\rho(t)$ is Gaussian for all $t$, and thus it can be fully characterized by the mean 
vector $\mean{x}=[\mean{q}, \mean{p}]^\top$ and the covariance matrix 
\[
     V = \left[\begin{array}{cc}
                  \mean{\Delta q^2} & \mean{\Delta q\Delta p + \Delta p\Delta q}/2 \\
                  \mean{\Delta q\Delta p + \Delta p\Delta q}/2  & \mean{\Delta p^2}  \\
           \end{array} \right], 
\]
where $\Delta q=q-\mean{q}$ and $\Delta p=p-\mean{p}$. 
These statistical variables are subjected to the equations $d\mean{x}/dt=A\mean{x}$ and 
$dV/dt=AV+VA^\top + D$, where 
\[
    A = \left[\begin{array}{cc}
               -2\sqrt{\kappa\gamma}-\gamma & 0  \\
               0 & -\gamma  \\
           \end{array} \right],~~
    D = \left[\begin{array}{cc}
               \gamma & 0  \\
               0 & (\sqrt{\kappa}+\sqrt{\gamma})^2  \\
           \end{array} \right].     
\]
The derivation of these matrices is given in Appendix~C. 
Then, in the limit $t\rightarrow \infty$, $\mean{x(t)}\rightarrow 0$ and 
$V(t)$ converges to the diagonal matrix 
${\rm diag}(\mean{\Delta q(\infty)^2}, \mean{\Delta p(\infty)^2})$ with 
\[
     \mean{\Delta q(\infty)^2}=\frac{\sqrt{\gamma}}{4\sqrt{\kappa}+2\sqrt{\gamma}},~~
     \mean{\Delta p(\infty)^2}=\frac{(\sqrt{\kappa}+\sqrt{\gamma})^2}{2\gamma}. 
\]
Clearly, $\mean{\Delta q(\infty)^2}<1/2$, hence the squeezed state is generated 
by the CF control. 
For example when $\kappa=9\gamma$, the variances are $\mean{\Delta q(\infty)^2}=1/14$ 
and $\mean{\Delta p(\infty)^2}=8$, which corresponds to about 8.5 dB squeezing. 
In this case the purity is only ${\rm Tr}(\rho(\infty)^2)=1/\sqrt{4{\rm det}(V(\infty))}\approx 0.66$, 
but this would not be a serious issue for the application to quantum metrology. 
\\

{\it Remark 3:} 
Let us consider the setup where the two system-probe couplings occur in a wrong 
order along the feedback loop; 
the dissipative coupling represented by $L_1=\sqrt{\gamma}a$ first occurs, 
and secondly the dispersive one $L_2=\sqrt{\kappa}q$ occurs. 
In this case, the Hamiltonian is calculated as 
$H = \sqrt{\kappa \gamma}(qp+pq)/2$. 
The coupling operator is the same as before, i.e., $L=L_1+L_2=\sqrt{\gamma}a+\sqrt{\kappa}q$. 
Then the system matrices characterizing this linear system are given by 
\[
      A=\left[ \begin{array}{cc}
               -\gamma & 0 \\
               0 & -\gamma - 2\sqrt{\kappa \gamma} \\
             \end{array} \right], ~~
      D=\left[ \begin{array}{cc}
               \gamma & 0 \\
               0 & (\sqrt{\kappa} + \sqrt{\gamma})^2 \\
             \end{array} \right]. 
\]
Then, the steady covariance matrix of the dynamics $dV/dt=AV+VA^\top + D$ is obtained as 
\[
     V(\infty)
           =\frac{1}{2}\left[ \begin{array}{cc}
                      1 & 0 \\
                      0 & 1+\kappa/(\gamma + 2\sqrt{\kappa \gamma}) \\
             \end{array} \right]. 
\]
Hence, the steady state is not a squeezed state. 
Note that, likewise the qubit case discussed in Remark~1, this results 
emphasizes the importance of the ordering of the two couplings.


\section{Fock state generation}

Lastly we consider the problem for generating a Fock state via feedback. 
The system is a high-Q optical cavity containing a few photons. 
In Refs.~\cite{Geremia 2006,Yanagisawa 2006,Molmer 2007}, the dispersive coupling 
$L_1=\sqrt{\kappa}n$, where $n=a^\dagger a$ with $a$ the annihilation operator of the 
cavity mode, was taken for MF control; 
this is the cross Kerr coupling between the cavity field and the probe field represented by 
$B_1$, the instantaneous Hamiltonian of which is given by 
\[
          H_{\rm int}^{(1)}(t+dt, t) 
               = i\sqrt{\kappa}\Big( n dB_1^\dagger (t) - n dB_1(t) \Big).
\]
In fact, this coupling induces a phase shift on $B_1$ depending on the number of photons 
inside the cavity; 
hence, by measuring the output field represented by $d\tilde{B}_1(t)=\sqrt{\kappa}j_t(n)dt+dB_1(t)$ 
(see Eq.~\eqref{output eq}), we can estimate the number of cavity photons and probabilistically 
obtain one of the eigenstates of $L_1$, i.e., a conditional Fock state $\ket{m}$.

Our aim is to construct a dissipative CF controller that compensates the dispersive 
process $L_1$ and produces a target Fock state deterministically. 
A simple dissipative process is the optical decay $L_2=\sqrt{\gamma}a$, represented by 
the interaction Hamiltonian 
\[
          H_{\rm int}^{(2)}(t+dt, t) 
               = i\sqrt{\gamma}\Big( a dB_2^\dagger (t) - a^\dagger dB_2(t) \Big), 
\]
where $B_2(t)$ is the annihilation field operator of the corresponding optical field. 
The CF control is structured by connecting the output $\tilde{B}_1$ to the input $B_2$.

Moreover, we add a displacement Hamiltonian $H_{\rm sys}=ig(a^\dagger - a)$, where 
$g$ is the gain to be determined, to move the steady state; 
note that merely the vacuum is produced if $H_{\rm sys}=0$. 
Also we take $\phi=0$. 
Hence, the CF controlled system \eqref{control ME} is characterized by 
\begin{equation}
\label{Fock L H}
       L = \sqrt{\kappa}n+\sqrt{\gamma}a, ~
       H = ig(a^\dagger - a) + \frac{\sqrt{\kappa\gamma}}{2i}(a^\dagger n - na). 
\end{equation}

\begin{figure}[htbp]
\begin{center}
\includegraphics[width=8.6cm]{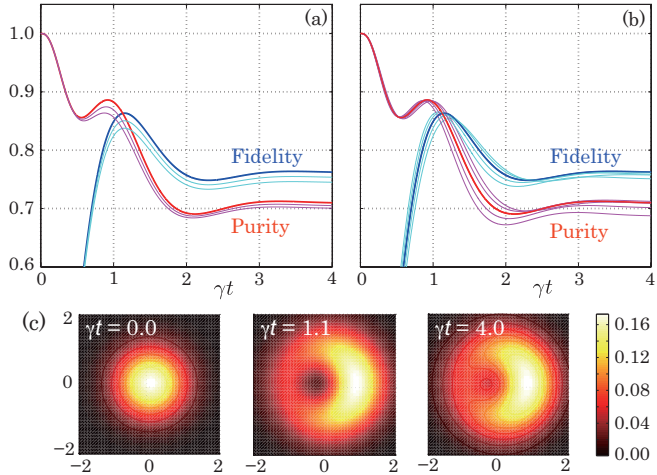}
\caption{
(a, b) Time evolution of the fidelity $F(t)=\bra{1}\rho(t)\ket{1}$ and the purity 
$P(t)={\rm Tr}(\rho(t)^2)$; 
the solid red and blue lines are the case of ideal setup, while cyan and magenta lines are 
the case under (a) decoherence and (b) parameter mismatch. 
(c) Q function of the system state at $\gamma t=0, 1.1, 4.0$ in the ideal setting. 
}
\label{fig:singlephoton}
\end{center}
\end{figure}

Now we fix the target to the single-photon $\ket{1}$, with initial state $\rho(0)=\ket{0}\bra{0}$. 
The control parameters are $\gamma=\kappa/4$ and $g=\kappa/2$, which are chosen 
to maximize the fidelity $F(t)=\bra{1}\rho(t)\ket{1}$, at some point of  time $t$. 
The blue and red lines in Fig.~\ref{fig:singlephoton}(a,b) show the time-evolution of 
$F(t)$ and the purity $P(t)={\rm Tr}(\rho(t)^2)$, respectively; 
the maximum fidelity is $F\approx 0.86$ at $\gamma t=1.1$ (with $P\approx 0.86$), and 
$F(t)$ converges to $F\approx 0.76$ (with $P\approx 0.71$). 
Therefore, the proposed CF controller actually emulates the MF scheme and generates 
a state close to $\ket{1}$ before reaching to the steady state which still has a feature of 
$\ket{1}$, as indicated by the Q function $Q(\alpha)=\bra{\alpha}\rho\ket{\alpha}/\pi$ 
shown in Fig.~\ref{fig:singlephoton}(c).

We also should study the effect of imperfection. 
In practice, there exists an uncontrollable photon leakage; 
we model this imperfection by introducing an extra optical field $B_3$ coupled to the cavity 
through the interaction Hamiltonian 
\[
       H_{\rm int}^{(3)}(t+dt, t) = i\sqrt{\epsilon}\Big( a dB_3^\dagger (t) - a^\dagger dB_3(t) \Big).
\]
The master equation of the CF-controlled system is then given by 
\[
       \frac{d\rho}{dt} = -i[H, \rho] + {\cal D}[L]\rho + {\cal D}[L_{\rm ex}]\rho,~~~
       L_{\rm ex}=\sqrt{\epsilon}a, 
\]
with $(L, H)$ given in Eq.~\eqref{Fock L H}. 
In Ref. \cite{Geremia 2006} the author estimated $\epsilon=12$ kHz while $\kappa=2.5$ MHz, 
which leads to $\epsilon\approx \kappa/200$; 
hence we take $\epsilon=\kappa/50, \kappa/100$. 
The cyan and magenta lines in Fig.~\ref{fig:singlephoton}(a) represent $F(t)$ and $P(t)$, 
respectively, in this imperfect setting. 
The figure shows that the peak fidelity $F(t)=\bra{1}\rho(t)\ket{1}$ at $\gamma t=1.1$ 
decreases from the optimal value 0.86 to 0.84. 
Apart from the decoherence, we have studied the case where the gain parameter 
$g$ in the displacement operation $ig(a^\dagger -a)$ deviates from the optimal value 
$g=\kappa/2$. 
Figure~6(b) shows the case $g=\kappa/2+\Delta$ with $\Delta=\pm \kappa/40, \pm\kappa/80$, 
while $\epsilon=0$ is assumed. 
Then from the figure we find that the fluctuation of the peak fidelity at $\gamma t=1.1$ is 
smaller than the case of decoherence. 
In summary, in both cases (a, b), the performance degradation is not so big, hence the CF 
scheme for the single photon generation is robust against those practical imperfections. 
This is in stark contrast to the MF strategy \cite{Haroche 2011} where $\ket{1}$ is generated 
with fidelity $F\approx 0.9$ but is collapsed immediately.


\section{Conclusion}

Ini this paper we demonstrated that a CF control can replace the MF one for the purpose 
of state preparation in some typical settings. 
The CF controller has a common structure, which is simply a series of dispersive and dissipative 
couplings inspired by the corresponding MF operation. 
Hence, it would have a wide-applicability in practice and work for other important objectives 
such as the quantum error correction. 
In fact, some studies along this direction have been conducted in a particular setup 
\cite{Mabuchi PRL 2010,Fujii 2014}. 
The sophisticated design theory for dissipative quantum networks \cite{James 2010} 
would be useful to solve those problems.

This work was supported in part by JSPS Grant-in-Aid No. 15K06151 and JST PRESTO 
No. JPMJPR166A. 
N.Y. acknowledges helpful discussions with M. Takeuchi.


\appendix


\section{Markovian open quantum systems}

\subsection{Quantum stochastic differential equation and master equation}

Here we derive the dynamical equation and the master equation of a general Markovian open 
quantum system that interacts with a single coherent field.

Let $b(t)$ be the annihilation operator of the coherent field and assume that $b(t)$ instantaneously 
interacts with the system. 
$b(t)$ satisfies the canonical commutation relation 
$[b(t), b^\dagger (s)]=\delta(t-s)$. 
As in the classical case, such a white noise process can be rigorously treated by introducing 
the annihilation process operator $B(t)=\int_0^t b(s)ds$; 
in particular, the infinitesimal change $dB(t)=B(t+dt)-B(t)$ satisfies the following 
{\it quantum Ito rule} \cite{Gardiner Book}: 
\begin{equation}
\label{Q ito rule}
     dtdB=0,~~ dBdB^\dagger=dt, ~~ dB^2=(dB^\dagger)^2=dB^\dagger dB=0. 
\end{equation}
The system-field interaction in the short time interval $[t, t+dt)$ is generally described by 
the Hamiltonian 
\begin{equation}
\label{int H}
      H_{\rm int}(t+dt, t) = i\Big( L dB^\dagger (t) - L^\dagger dB(t) \Big),
\end{equation}
where $L$ is a system operator representing the coupling with the field. 
The corresponding unitary operator in this time interval is given by 
$U(t+dt, t)={\rm exp}[-i H_{\rm int}(t+dt, t)]$. 
Then the total unitary operator from time $0$ to $t$, denoted by $U(t)$, is constructed by 
$U(t+dt)=U(t+dt,t) U(t)$, and from the quantum Ito rule \eqref{Q ito rule} we can 
derive the time evolution of $U(t)$ as follows: 
\begin{align}
& \hspace*{0.5em}
          U(t+dt)
\nonumber \\ & \hspace*{1em}
            = {\rm exp}[-i Hdt -i H_{\rm int}(t+dt, t)] U(t)
\nonumber \\ & \hspace*{1em}
            = \Big[ I -i Hdt -i H_{\rm int}(t+dt, t) - \frac{1}{2}H_{\rm int}(t+dt, t)^2 \Big] U(t)
\nonumber \\ & \hspace*{1em}
            = \Big[ I - (iH + \frac{1}{2}L^{\dagger}L)dt
                         + L dB^\dagger(t) - L^{\dagger} dB(t) \Big]U(t), 
\label{unitary QSDE}
\end{align}
with $U(0)=I$, where we have added the time-invariant system Hamiltonian $H$ (thus, 
the total Hamiltonian is $Hdt+H_{\rm int}(t+dt, t)$). 
From $dU(t)=U(t+dt)-U(t)$, Eq.~\eqref{unitary QSDE} is equivalently represented by 
\[
        dU(t) = \Big[ - ( iH + \frac{1}{2}L^{\dagger}L)dt
                         + L dB^\dagger(t) - L^{\dagger} dB(t) \Big]U(t),
\]
with $U(0)=I$. 
This is called the quantum stochastic differential equation (QSDE). 
Thus, a Markovian open quantum system $G$, which interacts with a single coherent field, is 
generally characterized by two operators $L$ and $H$, and thus it is denoted by $G=(L, H)$.

For an arbitrary system operator $X$, the Heisenberg equation of 
$X(t)=j_t(X)=U^{\dagger}(t)XU(t)$ is given by 
\begin{align}
& \hspace*{-0.5em}
    dX(t) = U^\dagger (t+dt)XU(t+dt) - U^\dagger (t)XU(t) 
\nonumber \\ & \hspace*{2.1em}
             = dU^\dagger (t) X U(t) + U^\dagger (t) X dU(t) + dU^\dagger (t) X dU(t)
\nonumber \\ & \hspace*{2.1em}
     = j_t\Big( i[H,X] + L^\dagger XL 
                -\frac{1}{2}L^\dagger LX - \frac{1}{2}XL^\dagger L \Big)dt
\nonumber \\ & \hspace*{4em}
 \mbox{}
        + j_t([X,L])dB^{\dagger}(t) + j_t([L^{\dagger},X])dB(t), 
\label{QSDE}
\end{align}
which is also called the QSDE. 
The field operator changes to $\tilde{B}(t)=j_t(B(t))$ and satisfies the output equation
\begin{equation}
\label{output eq}
          d\tilde{B}(t)=j_t(L)dt+dB(t). 
\end{equation}
Let us assume that the probe is a coherent field with amplitude $\alpha$. 
Then the expectation $\mean{X(t)}$ obeys 
\[
     \frac{d\mean{X(t)}}{dt}
         =\Big\langle j_t\Big( i[H',X] 
                + L^\dagger XL - \frac{1}{2}L^\dagger LX - \frac{1}{2}XL^\dagger L \Big) \Big\rangle,
\]
where $H'=H+(\alpha L^\dagger-\alpha^* L)/2i$. 
In the Schr\"{o}dinger picture the expectation $\mean{X(t)}$ is represented in terms of 
the time-dependent unconditional state $\rho(t)$ as $\mean{X(t)}=\mathrm{Tr}[X\rho(t)]$. 
Then it is easy to find that $\rho(t)$ obeys the master equation (1): 
\begin{equation}
\label{ME suppl}
       \frac{d\rho}{dt} = -i[H, \rho] + {\cal D}[L]\rho,~~
       {\cal D}[L]\rho = L \rho L^{\dagger}
                   - \frac {1}{2}L^{\dagger}L\rho - \frac {1}{2}\rho L^{\dagger}L, 
\end{equation}
where $H'$ has been replaced by $H$. 
Note finally that, if the system interacts with $m$ probe fields, then the resulting 
master equation is given by 
\begin{equation}
\label{general ME suppl}
       \frac{d\rho}{dt} = -i[H, \rho] + \sum_{k=1}^m{\cal D}[L_k]\rho. 
\end{equation}

\subsection{Derivation of the series product formula (2)}

The series product formula (2):
\begin{equation}
\label{cascade formula suppl}
        G_1\triangleright G_2 = \Big(L_1+L_2, ~H_1 + H_2 + 
              \frac{1}{2i}(L_2^\dagger L_1-L_1^\dagger L_2) \Big) 
\end{equation}
is directly obtained from Eq.~\eqref{unitary QSDE} as follows. 
Because the single probe field represented by $B(t)$ first interacts with the system 
$G_1=(L_1, H_1)$ and secondly with $G_2=(L_2, H_2)$, the change of the total 
unitary operator $U(t)$ is given by 
\begin{align}
& \hspace*{0.2em}
          U(t+dt) 
\nonumber \\ & \hspace*{0.3em}          
          = \Big[ I  - ( iH_2 + \frac{1}{2}L_2^{\dagger}L_2)dt
                         + L_2 dB^\dagger(t) - L_2^{\dagger} dB(t) \Big] 
\nonumber \\ & \hspace*{0.8em}
         \times  \Big[  I - ( iH_1 + \frac{1}{2}L_1^{\dagger}L_1)dt
                         + L_1 dB^\dagger(t) - L_1^{\dagger} dB(t) \Big] U(t)
\nonumber \\ & \hspace*{0.3em}
          = \Big[ I - i\Big( H_1 + H_2+\frac{1}{2i}(L_2^\dagger L_1-L_1^\dagger L_2) \Big)dt 
\nonumber \\ & \hspace*{2.2em}
             \mbox{}    - \frac{1}{2}(L_1+L_2)^{\dagger}(L_1+L_2)dt
\nonumber \\ & \hspace*{2.2em}
             \mbox{}  + (L_1+L_2) dB^\dagger(t) - (L_1+L_2)^{\dagger} dB(t) \Big] U(t). 
\nonumber
\end{align}
This means that the whole system $G_1\triangleright G_2$ is characterized by 
Eq.~(2) or \eqref{cascade formula suppl}. 
Note that, if $G_1$ and $G_2$ are different systems (for example, $G_1$ is a qubit and 
$G_2$ is an amplifier), then $(L_1, H_1)$ and $(L_2, H_2)$ are operators on the 
respective Hilbert spaces, and the more precise expression of the operators appearing 
in Eq.~\eqref{cascade formula suppl} is, e.g., $L_1\otimes I+I\otimes L_2$.

\subsection{The general SLH formula}

A more general Markovian open quantum system, which couples with $m$ independent 
probe fields $B(t)=[B_1(t), \ldots, B_m(t)]^\top$, is characterized by the triplet $(S, L, H)$, 
where $S$ is an $m\times m$ unitary matrix representing the scattering process of the 
probe fields. 
In this case the QSDE is represented by \cite{Gough 2009} 
\begin{align*}
& \hspace*{-1em}
          dU(t) = \Big[
                    - ( iH + \frac{1}{2}L^{\dagger}L)dt
                    + \mathrm{Tr}[(S-I)d\Lambda(t)^T] 
\nonumber \\ & \hspace*{5em}
                         + dB(t)^{\dagger}L-L^{\dagger}SdB(t) \Big] U(t),
\nonumber
\end{align*}
with $U(0)=I$, where $L=[L_1, \ldots, L_m]^\top$ is a vector of coupling operators, 
$\Lambda=(\Lambda_{ij})$ is the matrix of gauge process operators satisfying 
$d\Lambda_{ij}d\Lambda_{k\ell}=\delta_{jk}d\Lambda_{i\ell}$, 
and $H$ is a system Hamiltonian. 
It is shown in \cite{Gough 2009} that the cascade connection from $G_1=(S_1, L_1, H_1)$ 
to $G_2=(S_2, L_2, H_2)$ is given by 
\begin{align*}
& \hspace*{-1em}
    G_1\triangleright G_2 
        = \Big(S_2S_1,  ~ L_2+S_2L_1,  
\nonumber \\ & \hspace*{5em}
              ~  H_1 + H_2 + \frac{1}{2i}(L_2^\dagger S_2L_1-L_1^\dagger S_2^\dagger L_2) \Big). 
\nonumber
\end{align*}
The proposed CF controlled system (3) can then be equivalently represented by 
\[
       G=(1, L_1, H_{\rm sys}) \triangleright (e^{i\phi}, 0, 0) \triangleright (1, L_2, 0),
\]
where $(e^{i\phi}, 0, 0)$ represents a static device that only changes the phase of the field, 
such as a $\pi/2$ wave plate; 
that is, a phase shifter is placed along the feedback loop between the two systems, as 
shown in Fig.~\ref{fig:appendix} below.

\begin{figure}[htbp]
\begin{center}
\includegraphics[width=2.8cm]{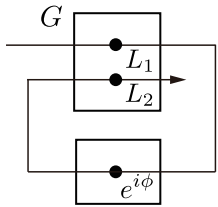}
\caption{
Coherent feedback configuration for the system $G$, composed of two couplings $L_1$ and 
$L_2$, and the phase shifter $e^{i\phi}$ placed along the feedback loop. }
\label{fig:appendix}
\end{center}
\end{figure}


\section{Proof of uniqueness of $\ket{\Phi_j}$ for the qutrit stabilization problem}

Here we prove that, in the ideal setup, one of the vectors $\{ \ket{\Phi_1}, \ket{\Phi_2}, \ket{\Phi_3} \}$ 
given in Sec.~IV-A can be selectively assigned as the unique pure steady state of the CF 
controlled system, by properly choosing the parameters $(u_1, u_2)$ in the added Hamiltonian 
$H_{\rm sys}$ given by Eq.~\eqref{qutrit Hsys}. 
First let us determine the parameter $(u_1, u_2)$, using Theorem~1 given in Sec.~III-A; 
that is, $\ket{\Phi_i}$ is a steady state of the master equation \eqref{eq:ME} if and only if it is 
an eigenvector of both $L$ and $iH+L^\dagger L/2$. 
Now $\{\ket{\Phi_1}, \ket{\Phi_2}, \ket{\Phi_3}\}$ are eigenvectors of $L$ in Eq.~\eqref{qutrit L H}. 
Then for $\ket{\Phi_1}$ to be a steady state, it must be an eigenvector of $iH+L^\dagger L/2$: 
\[
       iH+\frac{1}{2}L^\dagger L
        =  \left[\begin{array}{ccc}
                         (\kappa+\gamma)/2 & u_1 & 0 \\
                         -u_1 & \gamma/2 & u_2-\sqrt{\kappa\gamma}  \\
                         0 & -u_2 & \kappa/2  \\ 
                    \end{array} \right].
\]
That is, 
\begin{align}
& \hspace*{-0.5em}
        \left[\begin{array}{ccc}
                         (\kappa+\gamma)/2 & u_1 & 0 \\
                         -u_1 & \gamma/2 & u_2-\sqrt{\kappa\gamma}  \\
                         0 & -u_2 & \kappa/2  \\ 
                    \end{array} \right]
          \left[\begin{array}{c}
                         2\kappa  \\
                         2\sqrt{\kappa \gamma}  \\
                         \gamma \\ 
                    \end{array} \right]
\nonumber \\ & \hspace*{1em}
         = \left[\begin{array}{c}
                         \kappa(\kappa+\gamma)+2u_1\sqrt{\kappa\gamma}  \\
                         -2u_1 \kappa + u_2 \gamma  \\
                         -2u_2\sqrt{\kappa \gamma} + \kappa\gamma/2 \\ 
                    \end{array} \right]
           = \lambda 
           \left[\begin{array}{c}
                         2\kappa  \\
                         2\sqrt{\kappa \gamma}  \\
                         \gamma \\ 
                    \end{array} \right]
\nonumber
\end{align}
must hold, where $\lambda$ is an eigenvalue. 
This immediately yields $(u_1, u_2)=(-\sqrt{\kappa\gamma}/2, 0)$, with $\lambda=\kappa/2$. 
Similarly we obtain $(u_1, u_2)=(0, \sqrt{\kappa\gamma}/2)$ for the case $\ket{\Phi_2}$ 
and $u_2=\sqrt{\kappa\gamma}$ for the case $\ket{\Phi_3}$.

Now, by using the following result, we prove that $\ket{\Phi_i}$ is a {\it unique} steady state.

{\it Theorem 2} \cite{Kraus 2008}: 
Let ${\cal D}$ be the subset composed of pure steady states (called the ``dark states") of 
the Markovian master equation \eqref{general ME suppl} in the Hilbert space ${\cal H}$. 
If there is no subspace ${\cal S}\subseteq{\cal H}$ with ${\cal S}\perp{\cal D}$ such that 
$L_k{\cal S}\subseteq{\cal S}$ for all $k$, then ${\cal D}$ is the unique subset of steady states.

For the case $\ket{\Phi_1}$, ${\cal D}$ is given by ${\cal D}={\rm span}\{ \ket{\Phi_1} \}$. 
Then it is easy to find that the subspace orthogonal to ${\cal D}$ is 
\[
     {\cal S}={\rm span}\Big\{  
           \left[ \begin{array}{c}
               \gamma \\
               0 \\
               -2\kappa
             \end{array} \right], 
           \left[ \begin{array}{c}
               0 \\
               \sqrt{\gamma} \\
               -2\sqrt{\kappa}
             \end{array} \right] 
          \Big\}. 
\]
Then, we have 
\[
     L{\cal S}={\rm span}\Big\{  
           \left[ \begin{array}{c}
               \gamma \sqrt{\kappa}\\
               \gamma \sqrt{\gamma} \\
               2\kappa \sqrt{\kappa}
             \end{array} \right], 
           \left[ \begin{array}{c}
               0 \\
               0 \\
               1
             \end{array} \right] 
          \Big\}, 
\]
which clearly shows that $L{\cal S}\nsubseteq{\cal S}$. 
Therefore, from Theorem~2, $\ket{\Phi_1}$ is the unique steady state of the 
master equation of the system $(L, H)$ with $(u_1, u_2)=(-\sqrt{\kappa\gamma}/2, 0)$. 
Then, from the equivalency of the uniqueness of the steady state and the deterministic 
convergence to it for a finite dimensional Markovian quantum system \cite{Schirmer 2010}, 
we arrive at the conclusion that any initial state $\rho(0)$ converges to $\ket{\Phi_i}$. 
Similarly, we can prove the uniqueness of $\ket{\Phi_2}$ and $\ket{\Phi_3}$.


\section{Linear open quantum systems}

\subsection{General single-mode linear model}

Here we describe the QSDE of a general single-mode open harmonic oscillator that interacts 
with a single field; 
for a general system composed of multiple harmonic oscillators, see 
\cite{Wiseman Book,NY Book}. 
This system is generally characterized by the quadratic Hamiltonian 
\[
    H=\frac{1}{2}x^\top Gx 
       = \frac{1}{2}
          [q, p]\left[ \begin{array}{cc}
               g_1 &g_2 \\
               g_2 &g_3 \\
             \end{array} \right]
             \left[ \begin{array}{c}
               q \\
               p \\
             \end{array} \right], ~~~ g_i\in{\mathbb R}, 
\]
and the coupling operator $L=c_1q+c_2p$ ($c_1, c_2\in{\mathbb C}$), where $x=[q, p]^\top$ 
is the vector of canonical variables of the oscillator, satisfying $qp-pq=i$. 
Note that, from Eq.~\eqref{int H}, the oscillator couples with the field via the following 
interaction Hamiltonian:
\[
      H_{\rm int}(t+dt, t) = i(c_1q+c_2p) dB^\dagger (t) - i(c_1q+c_2p)^\dagger dB(t).
\]
Then the QSDEs \eqref{QSDE} of $q(t)=j_t(q)$ and $p(t)=j_t(p)$, for the system $(L,H)$ 
described above, are given by 
\begin{align}
& \hspace*{-0.5em}
     dq(t)=(g_2+{\rm Im}(c_1c_2^*))q(t)dt + g_3 p(t)dt 
\nonumber \\ & \hspace*{2.5em}
     -ic_2^*dB(t)+ic_2dB^\dagger(t), 
\nonumber \\ & \hspace*{-0.5em}
     dp(t)=-g_1q(t)dt - (g_2+{\rm Im}(c_1^*c_2))p(t)dt 
\nonumber \\ & \hspace*{2.5em}     
     + ic_1^*dB(t)-ic_1dB^\dagger(t).
\nonumber
\end{align}
These set of equations can be summarized as 
\begin{equation}
\label{1 D linear system}
      dx(t) = Ax(t) dt + i\Sigma[C^\top dB^\dagger(t) - C^\dagger dB(t)], 
\end{equation}
where $x(t)=[q(t),~p(t)]^\top$, 
\[
     A:=\Sigma[G+{\rm Im}(C^\dagger C)],~~
     C=[c_1,~c_2],~~
     \Sigma=\left[ \begin{array}{cc}
               0 &1 \\
               -1 &0 \\
             \end{array} \right]. 
\]
Also the output field operator \eqref{output eq} is expressed as 
\begin{equation}
\label{1 D linear output}
     d\tilde{B}(t) = Cx(t) dt + dB(t).
\end{equation}

Due to the linearity of Eq.~\eqref{1 D linear system}, the quantum state $\rho(t)$ is Gaussian 
for all $t$, if $\rho(0)$ is Gaussian. 
Then the system is fully characterized by the mean vector 
$\mean{x(t)}=[\mean{q(t)}, \mean{p(t)}]^\top$ and the covariance matrix 
\[
     V(t) = \left[ \begin{array}{cc}
                 \mean{\Delta q(t)^2} & \star \\
                  \mean{\Delta q(t)\Delta p(t)+\Delta p(t)\Delta q(t)}/2 & \mean{\Delta p(t)^2}  \\
             \end{array} \right], 
\]
%
%
where $\Delta q=q-\mean{q}$ and $\Delta p=p-\mean{p}$, and $\star$ denotes the symmetric 
element. 
The dynamics of $\mean{x(t)}$ is readily obtained as 
$d\mean{x(t)}/dt = A\mean{x(t)}$, where the field state is assumed to be the vacuum. 
Also from the quantum Ito rule \eqref{Q ito rule}, the time evolution equation of $V(t)$ is obtained as 
\begin{equation}
\label{lyapunov suppl}
    \frac{d}{dt}V(t) = AV(t)+V(t)A^\top + D,
\end{equation}
where $D=\Sigma{\rm Re}(C^\dagger C)\Sigma^\top$. 
It is known that, if all the eigenvalues of $A$ have negative real part, the mean vector 
$\mean{x(t)}$ converges to zero and Eq.~\eqref{lyapunov suppl} has a unique steady 
solution $V(\infty)$.

\subsection{Steady covariance matrix for the spin squeezing problem}

We here apply the above formulas to our model, and derive the dynamical equations of the 
system variables $(q(t), p(t))$ and the covariance matrix $V(t)$. 
Now the system is an open quantum harmonic oscillator driven by the following 
Hamiltonian and the coupling operator: 
\[
      H = -\frac{\sqrt{\kappa \gamma}}{2}(qp+pq), ~~
      L = (\sqrt{\kappa}+\sqrt{\gamma})q + i\sqrt{\gamma}p
\]
Hence, by definition we find 
\[
      G=\left[ \begin{array}{cc}
               0 & -\sqrt{\kappa \gamma} \\
               -\sqrt{\kappa \gamma} &0 \\
             \end{array} \right],~~
      C=[\sqrt{\kappa}+\sqrt{\gamma},~~i\sqrt{\gamma}]. 
\]
Then $A$ and $D$ in Eqs.~\eqref{1 D linear system} and \eqref{lyapunov suppl} are 
obtained as follows; 
\[
      A = \left[ \begin{array}{cc}
               -2\sqrt{\kappa\gamma}-\gamma & 0  \\
               0 & -\gamma \\
             \end{array} \right], ~~
      D =  \left[ \begin{array}{cc}
               \gamma & 0  \\
               0 & (\sqrt{\kappa}+\sqrt{\gamma})^2 \\
             \end{array} \right].
\]
Hence, the differential equation \eqref{lyapunov suppl} has the following unique 
steady solution: 
\[
      V(\infty)=\frac{1}{2}
            \left[ \begin{array}{cc}
                \sqrt{\gamma}/(2\sqrt{\kappa}+\sqrt{\gamma}) & 0  \\
               0 & (\sqrt{\kappa}+\sqrt{\gamma})^2/\gamma \\
             \end{array} \right].
\]
%


\end{document}